\documentclass{article}
\usepackage{spconf,amsmath,graphicx}

\usepackage{color}
\usepackage{multirow}
\usepackage{comment}
\usepackage{subfigure}
\usepackage{colortbl}

\newlength\savedwidth

\makeatletter
\def\bstctlcite{\@ifnextchar[{\@bstctlcite}{\@bstctlcite[@auxout]}}
\def\@bstctlcite[#1]#2{\@bsphack
\@for\@citeb:=#2\do{%
\edef\@citeb{\expandafter\@firstofone\@citeb}%
\if@filesw\immediate\write\csname #1\endcsname{\string\citation{\@citeb}}\fi}%
\@esphack}
\makeatother

\title{Sound Event Triage: Detecting Sound Events Considering Priority of Classes}
\name{Noriyuki Tonami$^{\dagger}$, Keisuke Imoto$^\diamondsuit$
}
\address{$^{\dagger}$\hspace{1pt}Ritsumeikan University, Japan, $^{\diamondsuit}$\hspace{1pt}Doshisha University, Japan}
\begin{document}
\ninept
\maketitle

\begin{abstract} % abstract
We propose a new task for sound event detection (SED): sound event triage (SET).
The goal of SET is to detect an arbitrary number of high-priority event classes while allowing misdetections of low-priority event classes where the priority is given for each event class.
In conventional methods of SED for targeting a specific sound event class, it is only possible to give priority to a single event class.
Moreover, the level of priority is not adjustable, i.e, the conventional methods can use only types of target event class such as one-hot vector, as inputs.
To flexibly control much information on the target event, the proposed SET exploits not only types of target sound but also the extent to which each target sound is detected with priority.
To implement the detection of events with priority, we propose class-weighted training, in which loss functions and the network are stochastically weighted by the priority parameter of each class.
As this is the first paper on SET, we particularly introduce an implementation of single target SET, which is a subtask of SET.
Results of the experiments using the URBAN--SED dataset show that the proposed method of single target SET outperforms the conventional SED method by 8.70, 6.66, and 6.09 percentage points for ``air\_conditioner,'' ``car\_horn,'' and ``street\_music,'' respectively, in terms of the intersection-based F-score.
For the average score of classes, the proposed methods increase the intersection-based F-score by up to 3.37 percentage points compared with the conventional SED and other target-class-conditioned models.
\end{abstract}


%%%%%%%%%%%%%%%%%%%%%%%%%%%%%%%%%%%%%%%%%%%%%%%%
%%                                            %%
%% The Main Body begins here                  %%
%%                                            %%
%% Please refer to the instructions for       %%
%% authors on:                                %%
%% https://www.biomedcentral.com/getpublished %%
%% and include the section headings           %%
%% accordingly for your article type.         %%
%%                                            %%
%% See the Results and Discussion section     %%
%% for details on how to create sub-sections  %%
%%                                            %%
%% use \cite{...} to cite references          %%
%%  \cite{koon} and                           %%
%%  \cite{oreg,khar,zvai,xjon,schn,pond}      %%
%%                                            %%
%%%%%%%%%%%%%%%%%%%%%%%%%%%%%%%%%%%%%%%%%%%%%%%%

%%%%%%%%%%%%%%%%%%%%%%%%% start of article main body
% <put your article body there>

%%%%%%%%%%%%%%%%%%%%%%%%
\section{Introduction}
%%%%%%%%%%%%%%%%%%%%%%%%

In our everyday life, humans utilize much information obtained from various environmental sounds \cite{Imoto_AST2018_01}.
The automatic analysis of environmental sounds will lead to the realization of many applications, e.g., anomalous sound detection systems \cite{koizumi_taslp2019_01}, life-logging systems \cite{Stork_ROMAN2012_01}, systems for hard-of-hearing persons \cite{Peng_ICME2009_01}, systems for smart cars \cite{Nandwana_INTERSPEECH2016_01}, and monitoring systems \cite{Ntalampiras_ICASSP2009_01}. 

Sound event detection (SED) \cite{Mesaros_SPmaga2021_01} is a major task in environmental sound analysis, which identifies sound event classes (e.g., ``dog barking,'' ``car passing by,'' and ``people walking'') with those time stamps.
In conventional SED, many methods using the hidden Markov model (HMM) \cite{Mesaros_eusipco2010_01,Heittola_JASM2013_01} and non-negative matrix factorization (NMF) \cite{Gemmeke_WASPAA2013_01,Komatsu_DCASE2016_01} have been proposed.
Recently, numerous deep neural network (DNN)-based SED methods have been in developed.
In DNN-based SED, the convolutional neural network (CNN) \cite{Hershey_ICASSP2017_01}, recurrent neural network (RNN) \cite{Hayashi_TASLP2017_01}, and convolutional bidirectional gated recurrent unit (CNN--BiGRU) \cite{cakir_TASLP2017_01} have been applied.
Moreover, some studies have shown that the self-attention-based Transformer \cite{kong_TASLP2020_01,miyazaki_icassp2020_01} and Conformer \cite{miyazaki_DCASEc2020_01} are useful for SED.

\begin{figure}[t!]
\centering
\includegraphics[width=0.9\columnwidth]{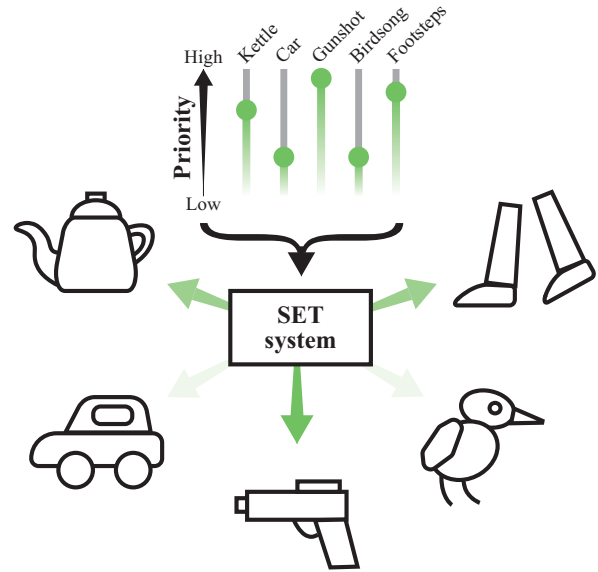}
\caption{Concept of SET task}
\label{fig:concept_SET}
\end{figure}

The target sounds to be analyzed depend on the user or application.
In the analysis of environmental sounds, a method to target a specific class in sound event localization and detection (SELD) has been proposed  \cite{Slizovskaia_arxiv2022_01}.
For SED, target sound detection (TSD) \textcolor{black}{has been proposed} by Yang \textit{et al.} \cite{Yang_arxiv2021_01}, in which only the single target sound class is detected, where a reference audio signal or a one-hot vector of the target sound is input to the TSD model.
In SED, the goal is to generalize the performance of detecting all sound events, i.e., the SED models are trained to equally detect all of the sound events in a dataset.
The widely used objective function of SED, binary cross entropy, is equally weighted for each event class.
In real environments, however, the detection priority for each event depends on the user or application.
For example, when SED is used for a surveillance system, anomalous events such as ``gunshot'' or ``baby crying'' have to be preferentially detected over other events.
On the other hand, in the case of a life-logging system, a sound event ``kettle'' or ``footsteps'' has to be more preferentially detected in addition to other events.
The conventional TSD system is trained only once on many event types and allows the user to choose a target event to focus on during inference. However, its limitation is that the degree of interest cannot be controlled.

To tackle this problem, we propose a new SED task: \textbf{sound event triage (SET)}.
The goal of SET is to improve the performance of detecting an arbitrary number of high-priority sound event classes while allowing the performance of detecting low-priority event classes to be compromised where the priority is given for each event class; that is, the triage.
In Fig. \ref{fig:concept_SET}, the concept of the SET task is illustrated.
A SET system enables user-preference sound event detection.
The difference between the conventional methods for SED including TSD \cite{Yang_arxiv2021_01} and the SET task is whether the degree to which events are detected with priority can be set.
Both SET and TSD models are trained once and select a class of interest.
Furthermore, only SET models can control the degree of the class of interest, i.e., the priority, at an inference stage.
For this first paper on SET, we design a network architecture for single target SET that is a subtask of SET, wherein the single event class is targeted with priority, and evaluate it in detail.
We propose a method for single target SET where loss-conditional training \cite{Dosovitskiy_iclr2020_01} is utilized for detecting sound events with priority.

%%%%%%%%%%%%%%%%%%%%%%%%
\section{Related works}
%%%%%%%%%%%%%%%%%%%%%%%%

\begin{figure*}[t!]
\centering
\includegraphics[width=2.0\columnwidth]{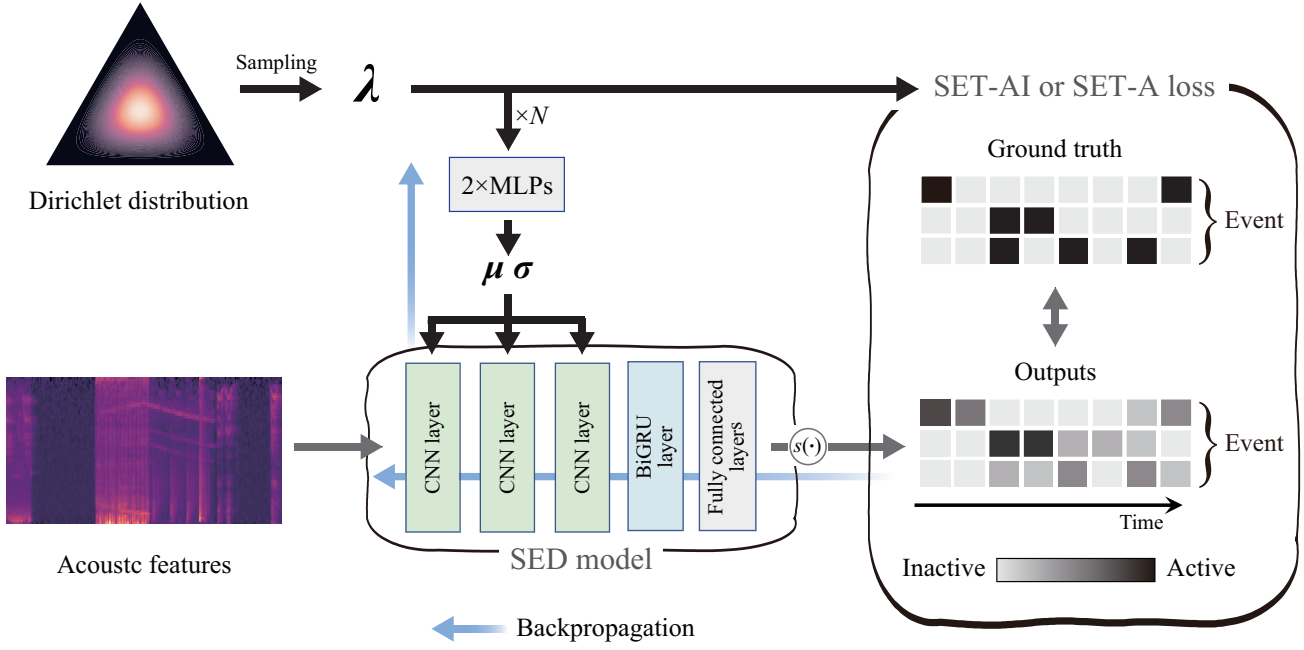}
\caption{Overview of SET in training stage}
\label{fig:overview_SET}
\end{figure*}

In this section, we describe works related to the proposed method.
In particular, the section comprises three subsections: strongly supervised SED, the conventional methods of environmental sound analysis using class-conditional techniques, and You Only Train Once (YOTO) \cite{Dosovitskiy_iclr2020_01}, with which the arbitrary linear combination of loss weights for multiple tasks can be set with a single model.

\subsection{Strongly supervised SED}
In strongly supervised SED, given a SED model $f$, model parameters $\pmb{\Theta}$, an acoustic feature ${\bf X}$, and a ground truth $z_{n,t} \in \{0,1\}$ for a sound event $n$ in time $t$, the SED model outputs the probability ${y}_{n,t} $ for the event $n$ and time frame $t$:

\begin{align}
    {y}_{n,t} = {\it P}( z_{n,t} \mid f, \pmb{\Theta}, {\bf X}).
    \label{eq:SED}
\end{align}

In the training of the DNN-based SED model, to optimize the model parameters $\pmb{\Theta}$, the following binary cross-entropy (BCE) loss function is used:

\begin{align}
{\mathcal L}_{\rm SED} 
 = - \sum^{N}_{n=1}& {\Big \{} {\bf z}_{n} \log {s}({\bf y}_{n}) 
\nonumber\\[0pt]
&\hspace{-10pt} + (1 - {\bf z}_{n}) \log {\big (} 1 - {s}({\bf y}_{n}) {\big )} {\Big \}}\\[5pt]
 = - \sum^{N}_{n=1}& \sum^{T}_{t=1}{\Big \{} z_{n,t} \log {s}(y_{n,t}) 
\nonumber\\[0pt]
&\hspace{-10pt} + (1 - z_{n,t}) \log {\big (} 1 - {s}(y_{n,t}) {\big )} {\Big \}},
\label{eq:event_loss}
\end{align}

\noindent where $s(\cdot)$ denotes the sigmoid function.
$N$ and $T$ are the numbers of sound event classes and total time frames, respectively.
In an inference stage of SED, $s(y_{n,t})$ is binarized with a predefined threshold to obtain detection results.
As can be seen in Eq. \ref{eq:event_loss}, all sound events are equally weighted for generalizing the performance of detecting all sound events.

\subsection{Methods of environmental sound analysis for targeting a specific sound}
In the analysis of environmental sounds, several methods considering user-guided or target-class-conditioned information have been proposed \cite{Slizovskaia_arxiv2022_01,Yang_arxiv2021_01}.
In works related to SED, TSD \cite{Yang_arxiv2021_01} and class-conditioned SELD where information on target event classes or sounds is employed \cite{Slizovskaia_arxiv2022_01} have been studied.
Yang \textit{et al}. \cite{Yang_arxiv2021_01} have proposed the TSD task derived from SED, which detects target sound events with a one-hot vector of the target event class.
In the TSD network, a reference signal or a one-hot vector of the target event class is input to a network \textcolor{black}{as a condition}, which is embedded and then fused with a SED network.
Slizovskaia \textit{et al}. \cite{Slizovskaia_arxiv2022_01} have proposed the class-conditioned SELD, which analyzes a specific-target event similarly to TSD to detect the target sound using a one-hot vector of the target class.
These conventional systems related to SED can only handle information about the types of sound to be analyzed\textcolor{black}{, not the degree of priority given to the events of special interest.}

\subsection{YOTO}
YOTO \cite{Dosovitskiy_iclr2020_01} is a technique that enables a single network to change into various specialist models without retraining in inference stages.
Each specialist model has better performance for a particular task.
For example, we assume two specialist models, image-quality and compression-rate specialists \cite{Dosovitskiy_iclr2020_01}.
In this case, a single network using YOTO can change into the two specialists or a model with an intermediate expertise of the two specialists without retraining in inference stages.
The YOTO scheme is efficient in terms of the training cost or model complexity compared with multiple networks for each type of expertise.

In YOTO, we assume a problem setting that a single DNN-based network performs multiple tasks where the network is trained with multiple losses for each task.
Let $\mathcal{L}_m$ be a loss function for task $m$.
The following loss function is often used for optimizing the parameters of the network:

\begin{align}
{\mathcal L} = \sum_{m=1}^{M} \lambda_{m} \mathcal{L}_{m}, 
\label{eq:multiple_task}
\end{align}
  
\noindent where $\lambda_{m} (0 \leq \lambda_{m} \leq 1.0 )$ is the $m$-th element of a vector {$\pmb \lambda$} for balancing among the tasks..
$M$ denotes the number of tasks.
In the training stage, the single network is trained with various $\pmb \lambda$.
In inference stages, an arbitrary $\pmb \lambda$ can be input to the single network trained.
When $\lambda_{m}$ is set to be larger than those in other tasks, the network focuses on the training and/or inference of task $m$ instead of other tasks.

%
%
%
%
%%%%%%%%%%%%%%%%%%%%%%%%
\section{Proposed method}
%%%%%%%%%%%%%%%%%%%%%%%%
%%%%%%%%%%%%%%%%%%%%%%%%
\subsection{Framework of SET}
%%%%%%%%%%%%%%%%%%%%%%%%
In the SET task, an arbitrary number of event classes are detected with priority.
In the training stage of SET, triage weights are given for detecting target events with priority in addition to acoustic features and model parameters.

\begin{align}
    {y}_{n,t} = {\it P}( z_{n,t} \mid f, \pmb{\Theta}, {\bf X}, \pmb{\lambda}),
    \label{eq:formula_SET}
\end{align}

\noindent where $\pmb{\lambda} = (\lambda_{1}, \lambda_{2}, ..., \lambda_{N})$ are the parameters for the triage, that is, detecting sound events with priority.
$\lambda_{n} (0 \leq \lambda_{n} \leq 1.0 )$ is a triage weight for the sound event $n$.
In an inference stage of SET, an arbitrary triage parameter $\pmb{\lambda}$ is input to the SET model. 
When $\lambda_{n}$ is set to a larger value than the others, the model targets with higher priority the sound event $n$ than the other events.
Examples of the loss functions on the right side in Fig. \ref{fig:overview_SET} are described in section 3.4.

\begin{figure}[t!]
\centering
\includegraphics[width=0.9\columnwidth]{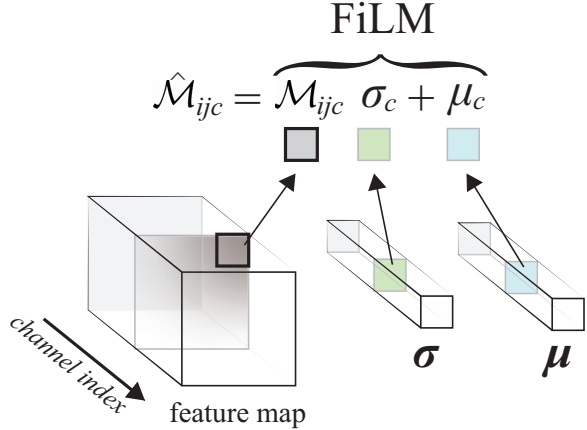}
\caption{Illustration of FiLM operation}
\label{fig:FiLM}
\end{figure}

%%%%%%%%%%%%%%%%%%%%%%%%
\subsection{Class-weighted training}
%%%%%%%%%%%%%%%%%%%%%%%%

In the DNN-based SED, as shown in Eq. \ref{eq:event_loss}, the BCE loss can be divided into losses of each event class.
In other words, the BCE loss can be regarded as the sum of losses for the task of detecting each sound event class.
To train a SET model, the following loss function is used:

\begin{align}
  {\mathcal L}_{\rm SET} &= {\mathcal L}(F({\bf X},{\pmb\lambda}),{\bf Z},{\pmb\lambda}) \\ 
  &= \sum_{n=1}^{N} N\lambda_{n} {\mathcal L}_{\rm SED}({\bf y_{\it n},z_{\it n}}) ,
  \label{eq:SET_loss}
\end{align}
where {\it F}($\cdot$) is the output of a SED model.
${\bf Z}$ and ${\pmb\lambda}$ indicate the outputs of the model and triage parameters, respectively.
${\mathcal L}_{\rm SED}({\bf y_{\it n},z_{\it n}})$ and $\lambda_{n}$ are the loss function and triage parameter for event class $n$, where ${\bf y_{\it n}}$ and ${\bf z_{\it n}}$ are $\{y_{n,1} \dots y_{n,t} \dots y_{n,T}\}$ and $\{z_{n,1} \dots z_{n,t} \dots z_{n,T}\}$, respectively.
\textcolor{black}{The loss functions} for each event class are weighted by the priority parameters of each class.
We call the training scheme using the loss function ``class-weighted training.''
In Eq. \ref{eq:SET_loss}, \textcolor{black}{we normalize $\pmb\lambda$ so that $\sum_n \lambda_n$ = $1.0$, and scale the loss function by multiplying by $N$.}
As can be seen from Eq. 6, both the SED model and the loss function are conditioned with the triage parameters.

To use arbitrary $\lambda_{n}$ in inference stages of a SET network, $\pmb\lambda = (\lambda_{1}, \lambda_{2},... ,\lambda_{N})$ are repeatedly and randomly sampled from a distribution during the training, which cover various $\pmb\lambda$ values in the single SET network.
The sampled parameters $\pmb{\lambda}$ are input to the SET network and used for the loss calculation (Eq. \ref{eq:SET_loss}) in the training stage.
As shown in  Fig. \ref{fig:overview_SET}, $\pmb \lambda$ values are firstly fed to two multilayer perceptrons (MLPs).
The MLPs output two vectors, $\pmb{\mu} = (\mu_{1}, \mu_{2},... ,\mu_{C})$ and $\pmb{\sigma} = (\sigma_{1}, \sigma_{2},... ,\sigma_{C})$.
As shown in Fig. \ref{fig:FiLM}, feature-wise linear modulation (FiLM) \cite{Perez_aaai2018_01} is then used to bridge between the outputs of the MLPs and the SED model for detecting events (Fig. \ref{fig:overview_SET}).
The FiLM is applied to a feature map in CNN layers of the SED model.
The feature map is multiplied by $\pmb \sigma$ and added to $\pmb \mu$: ${ \mathcal{\hat M}}_{ijc} = \mathcal{M}_{ijc}\sigma_{c}+\mu_{c}$.
Here, $\mathcal{M}_{ijc}$ is a feature at a location $(i,j)$ of a channel index $c$ in a feature map of a CNN layer, as shown in Fig. \ref{fig:FiLM}.
\textcolor{black}{FiLM has been reported to perform better than including conditional information, e.g., the triage weights, as additional inputs to the network \cite{Dosovitskiy_iclr2020_01}. 
Because all the CNN layers in our network have the same number of channels, we feed the same $\pmb{\mu}$ and $\pmb{\sigma}$ to all of them for convenience.}
In addition to the conditioning of the SED model, the sampled $\pmb \lambda$ values are directly used for the losses (Eq. \ref{eq:SET_loss}) in training stages.

\subsection{Single target SET}
As the initial work of SET, we introduce a model training scheme for the single target SET, which is a subtask of SET.
In Fig. \ref{fig:overview_SET}, the overview of single target SET method in the training stage is shown.
For single target SET, as the distribution of the triage weight $\lambda_{n}$, we use the Dirichlet distribution ${\mathcal D}(\pmb\alpha)$.
The probability density function of the $(K-1)$-dimensional Dirichlet distribution is

\begin{align}
{\mathcal D}(\pmb\alpha) = \frac{\Gamma(\sum_{k=1}^K\alpha_{k} )}{\prod_{k=1}^K\Gamma(\alpha_{k})}\prod_{k=1}^{K}x_{k}^{\alpha_{k}-1},\\
s.t. \sum_{k=1}^{K}x_{k}=1,
\label{eq:Dirichlet}
\end{align}

\noindent where $x_{k} \geq 0$ and $\alpha_{k} > 0$ are stochastic variables for a dimension $k$ and a parameter for the shape of the distribution.
$\Gamma(\cdot)$ represent the gamma function.
In a training stage of a SET model, we need a distribution that can be controlled to give a larger weight into a specific class (target) than into other classes (nontarget).
The Dirichlet distribution with smaller $\pmb \alpha$ produces such a vector for giving a larger weight into a specific class.
The Dirichlet distribution is suitable for single target SET because one class has a larger weight than the others by setting $\pmb\alpha$ smaller.
Our single target SET model is conditioned by class-weighted training using YOTO, where the triage weight $\pmb\lambda$ is sampled for each sound event class, to handle arbitrary $\pmb\lambda$, i.e., the priority of the detection.
$N$ is also multiplied by inputs of the MLPs for conditioning the SED model, as shown in Fig. \ref{fig:overview_SET}.

%%%%%%%%%%%%%%%%%%%%%%%%
\subsection{SET losses with priority of event classes}
%%%%%%%%%%%%%%%%%%%%%%%%
To perform our SET, Eq. \ref{eq:SET_loss} is specified in this section. 
We propose two loss functions for the class-weighted training of SET.
First, we introduce a loss function of SET with active and inactive frames (SET--AI) as follows.

\begin{align}
{\mathcal L}_{\rm SET\mathchar`-AI}
 = - \sum^{N}_{n=1}& N\lambda_{n}{\Big \{} {\bf z}_{n} \log {s}({\bf y}_{n}) 
\nonumber\\[0pt]
+& (1 - {\bf z}_{n}) \log {\big (} 1 - {s}({\bf y}_{n}) {\big )} {\Big \}}\\[5pt]
 = - \sum^{N}_{n=1}& \sum^{T}_{t=1}N\lambda_{n}{\Big \{} z_{n,t} \log {s}(y_{n,t}) 
\nonumber\\[0pt]
+ (1 &- z_{n,t}) \log {\big (} 1 - {s}(y_{n,t}) {\big )} {\Big \}}
\label{eq:SET_active_inactive}
\end{align}

\noindent In SET--AI, the triage weight $\lambda_{n}$ affects the active and inactive frames of sound event $n$.
When $\lambda_{n}$ is set to a larger value than the others, the model focuses on the training and/or inference of the active and inactive frames of event $n$ compared with the others.
In SET--AI, the loss of inactive frames is multiplied by $\lambda_{n}$.

\begin{figure}[t!]
\centering
\includegraphics[width=0.95\columnwidth]{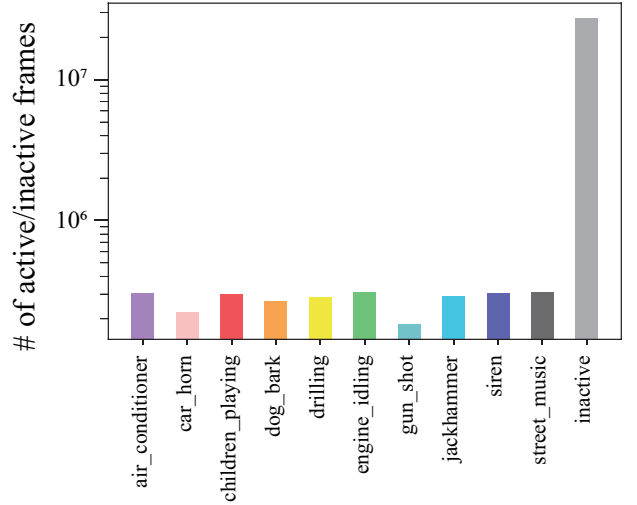}
\caption{Numbers of active and inactive frames for each event}
\label{fig:frames}
\end{figure}

A large number of inactive frames disturb the training of active frames, as reported in \cite{imoto_icassp2021_01}.
Hence, only the loss of active frames is multiplied by $\lambda_{n}$.
Thus, we also introduce a loss function of SET with active frames (SET--A), wherein the model focuses on the training of the active frames, as follows.

\begin{align}
{\mathcal L}_{\rm SET\mathchar`-A} 
 = - \sum^{N}_{n=1}& {\Big \{} N\lambda_{n}{\bf z}_{n} \log {s}({\bf y}_{n}) 
\nonumber\\[0pt]
+& (1 - {\bf z}_{n}) \log {\big (} 1 - {s}({\bf y}_{n}) {\big )} {\Big \}}\\[5pt]
 = - \sum^{N}_{n=1}& \sum^{T}_{t=1}{\Big \{} N\lambda_{n}z_{n,t} \log {s}(y_{n,t}) 
\nonumber\\[0pt]
+ (1 &- z_{n,t}) \log {\big (} 1 - {s}(y_{n,t}) {\big )} {\Big \}}
\label{eq:SET_active_only}
\end{align}

\noindent The difference between SET--AI and SET--A is whether the number of inactive frames is multiplied by the triage weight $\lambda_{n}$.

%%%%%%%%%%%%%%%%%%%%%%%%
\vspace{15pt}
\section{Experiments}
%%%%%%%%%%%%%%%%%%%%%%%%

%%%%%%%%%%%%%%%%%%%%%%%%
\subsection{Experimental conditions}
%%%%%%%%%%%%%%%%%%%%%%%%
To evaluate the effectiveness of our methods, we conducted the following experiments:

\begin{itemize}
    \setlength{\itemsep}{5pt}
    \item[] \textbf{[Experiment 1]}: We verified that class-weighted training enables the detection of sound events with priority in terms of F-scores (sections 4.2.1 and 4.2.2).
    \item[] \textbf{[Experiment 2]}: To analyze in more detail the properties of the proposed methods, we observed misdetection results in terms of insertion or deletion rate for the proposed methods (section 4.2.3).
    \item[] \textbf{[Experiment 3]}: We investigated how the triage weights are affected for each event class and evaluation metric (section 4.2.4).
\end{itemize}

\vspace{5pt}
\begin{table}[t!]
	\small
	\caption{Experimental conditions}
	\vspace{0pt}
	\label{tbl:parameter}
	\centering
	\scalebox{0.95}[0.95]{
		\begin{tabular}{ll}
			\hline
			&\\[-9pt]
			\textbf{SED model}\\
			& \\[-9pt]
			Network architecture & 3 CNN + 1 BiGRU + 2 FC\\
			\# channels of CNN layers & 64, 64, 64 \\
			Filter size ($T\times F$)& 3$\times$3 \\
			Pooling size ($T\times F$)& 8$\times$1, 2$\times$1, 2$\times$1 (max pooling) \\
			&\\[-9pt]
			\# of units in BiGRU layer & 64 \\
			\# of units in FC layers & 32\\
			\# of units in output layer & 10 \\\hline
			&\\[-9pt]
			\textbf{MLPs for each $\pmb\mu$ and $\pmb\sigma$}\\
			& \\[-9pt]
		    Network architecture & 3 FC\\
			\# of units in FC layers & 64, 256, 128\\
			\# of units in output layer & 64 \\\hline
			&\\[-9pt]
			Optimizer & Adam \cite{Kingma_ICLR2015_01}\\
			Activation functions & leaky ReLU\\\hline
		\end{tabular}
	}
	\vspace{30pt}
\end{table}

\begin{figure*}[t!]
\centering
\includegraphics[width=2.0\columnwidth]{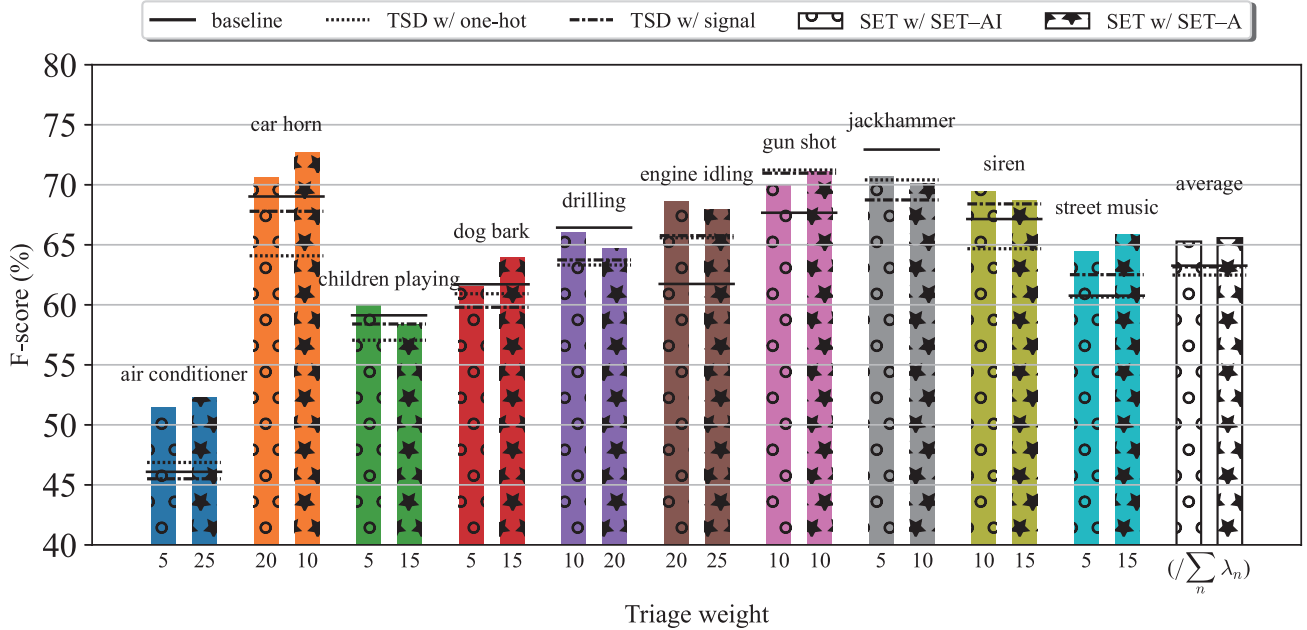}
\caption{\textcolor{black}{SET results in terms of frame-based F-score (\%) for target classes}}
\label{fig:result_01}
\end{figure*}

For the experiments, we used the URBAN--SED \cite{Salamon_WASPAA2017_01} dataset.
URBAN--SED includes 10,000 synthetic audio clips (train, 6,000; validation, 2,000; test, 2,000), where the duration of each clip is 10 s with a sampling rate of 44,100 Hz.
The dataset consists of 10 sound event classes.
In Fig. \ref{fig:frames}, the numbers of active and inactive frames for each event are indicated.
As acoustic features, we used 64-dimensional log-mel band energies, which were calculated with the window size of 40 ms and the hop size of 20 ms.
This setup is based on the baseline system of DCASE2018 Challenge task4 \cite{Serizel_DCASE2018_01}.
The threshold value for detecting events is tuned using the \textcolor{black}{validation set} for each event class and method \textcolor{black}{with the intersection-based F-score.}
\textcolor{black}{The other hyperparameters are also optimized with the intersection-based F-score.}
For post-processing before detection, a median filter is applied, where the filter size is tuned with the \textcolor{black}{validation set} for each event class and method.
The batch size was 64, and models were trained with 100 epochs.
To measure the detection performance, we used frame-based and intersection-based metrics  \cite{Bilen_ICASSP2020_01}.
In the intersection-based metric, the detection tolerance criterion (DTC) and ground truth intersection criterion (GTC) are both set to 0.5.

As the SED model in Fig. \ref{fig:overview_SET}, we used two models.
We \textcolor{black}{used} CNN--BiGRU with selective kernel units (CNN--BiGRU--SK) \cite{Zheng_ICASSP2021_01,Zheng_DCASEc2021_01}, which achieved the best performance in DCASE2021 Challenge task4.
In CNN--BiGRU--SK, kernels of multiple sizes are adopted in a CNN of a single model to handle various types of sound event.
Moreover, for comparison, we used the TSD model \cite{Yang_arxiv2021_01}.
We used two versions of the TSD model: conditioned with a one-hot vector and with a reference signal.
For the one-hot-vector-based TSD model, to make the conditions for TSD and our SET the same, we used only the detection and conditional networks.
In other words, we did not employ the classification loss, which is used only when the TSD model is conditioned by a reference signal.
We used the same architecture as that in the one-hot-vector-based TSD model for the detection network.
For the conditional network of the one-hot-vector-based TSD model, we used \textcolor{black}{the same two MLP layers as that in the single target SET method} because one-hot vectors of sound events were utilized instead of reference signal.
For the reference-signal-based TSD model, we randomly chose clips of UrbanSound8k \cite{Salamon_ACMMM14_01}, as in \cite{Yang_arxiv2021_01}.
Other detailed parameters are shown in Table \ref{tbl:parameter}.
In Table \ref{tbl:parameter}, ``FC'' means fully connected.

The FiLM operation (${\hat m}_{ijc} = \sigma_{c}m_{ijc}+\mu_{c}$) was implemented between convolution and max pooling in each CNN layer.
As shown in Fig. \ref{fig:overview_SET}, the SED model and two MLPs are simultaneously optimized using the loss function Eq. \ref{eq:SET_active_inactive} or \ref{eq:SET_active_only}.
In this work, $\alpha_{\forall k}$ is set to 0.1 for the symmetric Dirichlet distribution ${\mathcal D}(\pmb\alpha)$, which was tuned using the \textcolor{black}{validation set}.
$\pmb\lambda \sim {\mathcal D}(\pmb\alpha)$ is sampled for each batch of an epoch during the training of a SET model.
$K$ is set to the number of sound event classes.

%%%%%%%%%%%%%%%%%%%%%%%%
\vspace{55pt}
\subsection{Experimental results}
%%%%%%%%%%%%%%%%%%%%%%%%

\begin{figure}[t!]
\centering
\vspace{20pt}
\includegraphics[width=0.95\columnwidth]{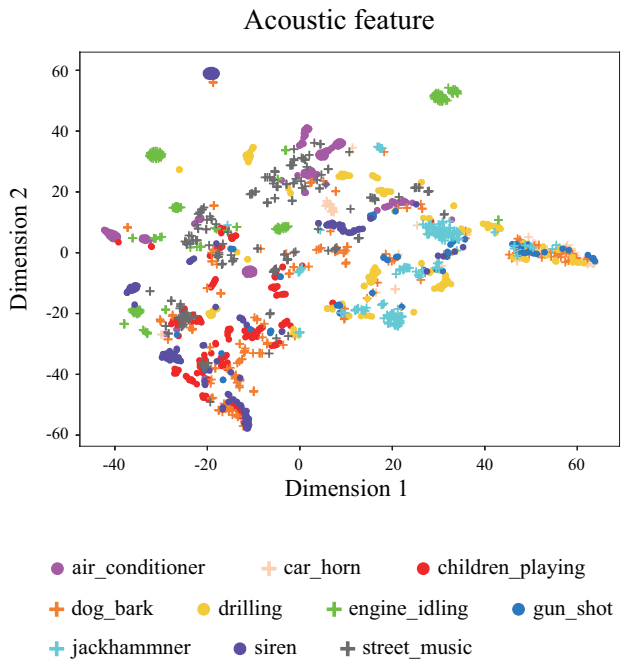}
\caption{Relationship among event classes on test set of UrbanSound8k in terms of acoustic features visualized using t-SNE}
\label{fig:visualization_acoustic_feature_space}
\vspace{15pt}
\end{figure}

\subsubsection{\rm\textbf{[Experiment 1]}: SET results in terms of frame-based F-score}
\vspace{5pt}
In this experiment, we selected one target event class $n$ for single target SET and then observed the F-scores of the target event class with various $\lambda_{n}$ values in inference stages.
The triage weight for a nontarget class is fixed at $1.0/\sum_{n}\lambda_n$.
For example, when the index of a target class is $n=1$ and the triage weight $\lambda_{1}$ is set to $5.0/\sum_{n}\lambda_n$, $\pmb\lambda = (5.0, 1.0, \ldots, 1.0)/\sum_{n}\lambda_n$.
In all of the experimental results, all detection results of our SET are obtained with the optimal triage weight tuned using the \textcolor{black}{validation set}.
The optimal triage weights are set for each method, class, and evaluation metric using the \textcolor{black}{validation set}.
As aforementioned in section 3.2, \textcolor{black}{$\pmb\lambda$ is multiplied by $N$} for the scaling before $\pmb\lambda$ is input to the two MLPs for $\pmb\mu$ and $\pmb\sigma$.

Figure~\ref{fig:result_01} shows the results of the proposed methods in terms of the frame-based F-score.
``baseline'' indicates the results of CNN--BiGRU--SK.
``TSD w/ one-hot'' and ``TSD w/ signal'' represent the TSD model conditioned by the one-hot vector and by the reference signal, respectively.
``SET w/ SET--AI'' and ``SET w/ SET--A'' are SET with class-weighted training using Eqs. \ref{eq:SET_active_inactive} and \ref{eq:SET_active_only}, respectively.
The results show that the proposed SET methods with class-weighted training achieved a reasonable performance.
For the average detection performance of the classes, the SET model with SET--A loss improved the frame-based F-score by 2.29 percentage points compared with the baseline value.
Moreover, the performance of detecting those sound events  increases when using the SET--A loss compared with using the SET--AI loss.
This is because the number of inactive frames of the sound events is large in the training set, as can be seen in Fig. \ref{fig:frames}.
In other words, the models using the SET--AI loss might focus on the training of inactive frames.
This leads to the degradation of the detection performance, as reported in \cite{imoto_icassp2021_01}.

\textcolor{black}{For individual classes,} the SET--A loss improved the performance of detecting the sound events ``air\_conditioner,'' ``car\_horn,'' and ``street\_music'' by 6.17, 3.65, and 5.12 percentage points, respectively, compared with the baseline values.
In particular, the detection performance of ``air\_conditioner'' using the conventional methods is lower than those of the other classes, but it markedly increased when using the proposed methods.
This indicates that our SET can boost the performance of recognizing the laborious-to-detect classes.
On the other hand, the sound events ``drilling'' and ``jackhammer'' are not well detected using the proposed SET compared with the baseline values.
This might be because ``drilling'' is acoustically similar to ``jackhammer.''
In \cite{Salamon_ACMMM14_01}, the timbre of these two events is similar and could also be confused in the classification task.
To confirm the similarities among the sound events, \textcolor{black}{we visualized the acoustic feature space} using t-SNE in Fig. \ref{fig:visualization_acoustic_feature_space}.
Note that we used UrbanSound8k \cite{Salamon_ACMMM14_01}, which is composed of isolated sound events of the URBAN--SED dataset for visualizing relationships among the sound event classes.
This is because it is difficult to clearly visualize the relationships owing to overlapped sound events, i.e., polyphonic, in an audio clip of the URBAN--SED dataset.
Figure \ref{fig:visualization_acoustic_feature_space} also indicates that ``drilling'' is acoustically similar to ``jackhammer.''
The SET--A loss, which focuses on the active frames, may detect a target sound event and similar one simultaneously compared with the SET--AI loss.

\begin{figure*}[t!]
\centering
\includegraphics[width=2.0\columnwidth]{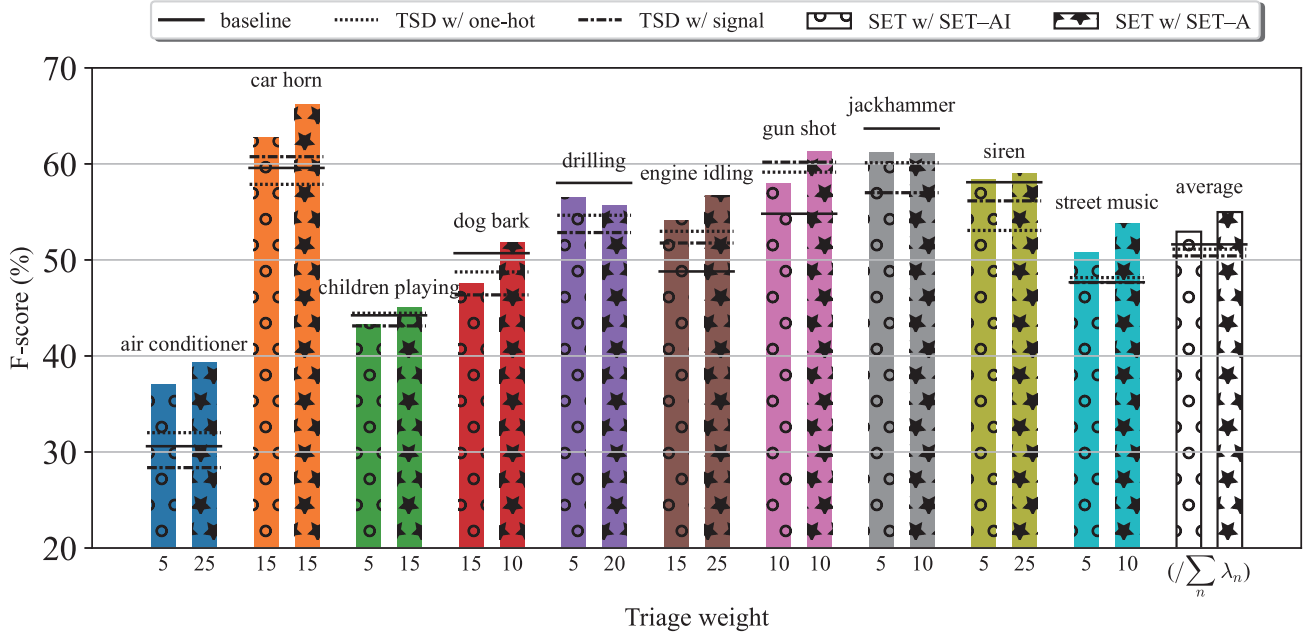}
\caption{SET results in terms of intersection-based F-score (\%) for target classes}
\label{fig:result_02}
\end{figure*}

We next compare the TSD models with our single target SET models.
As shown in Fig. \ref{fig:result_01}, our single target SET models detected many events better than the TSD model, e.g., ``air\_conditioner,'' ``car\_horn,'' and ``street\_music.''
The TSD models outperform the SED model for some events such as ``engine\_idling'' and ``gun\_shot.''
In particular, the F-score of ``gun\_shot'' when using the TSD models achieved a reasonable performance comparable to that of our single target SET.
As previously indicated in Fig. \ref{fig:frames}, the number of active frames of ``gun\_shot'' is smaller than those of the other classes.
This indicates that the targeting methods, such as TSD and SET, are useful for event classes where the number of frames is small, e.g., rare sound event classes.

\subsubsection{\rm\textbf{[Experiment 1]}: SET results in terms of intersection-based F-score}
\vspace{5pt}
We also evaluated the proposed methods in terms of the intersection-based F-score.
In the intersection-based F-score, unlike the frame-based F-score, models are evaluated instance by instance.
Here, ``instance'' means a block with the associated onset and offset \cite{Mesaros_MDPI2016_01}. 
Figure \ref{fig:result_02} is SET results in terms of the intersection-based F-score.
The results show that the F-score of the proposed SET method is improved compared with that of the baseline system.
For the average detection performance of the classes, the SET model with the SET--A loss increased the intersection-based F-score by 3.37 percentage points compared with the baseline model.
For each class, the SET--A loss improved the performance of detecting the sound events ``air\_conditioner,'' ``car\_horn,'' and ``street\_music'' by 8.70, 6.66, and 6.09 percentage points, respectively, compared with the baseline values.
Comparing Figs. \ref{fig:result_01} and \ref{fig:result_02}, we find that SET w/ SET--A greatly outperformed SET w/ SET--AI for most of the classes in terms of the intersection-based F-score rather than the frame-based F-score.

\begin{figure*}[t!]
\centering
\includegraphics[width=2.0\columnwidth]{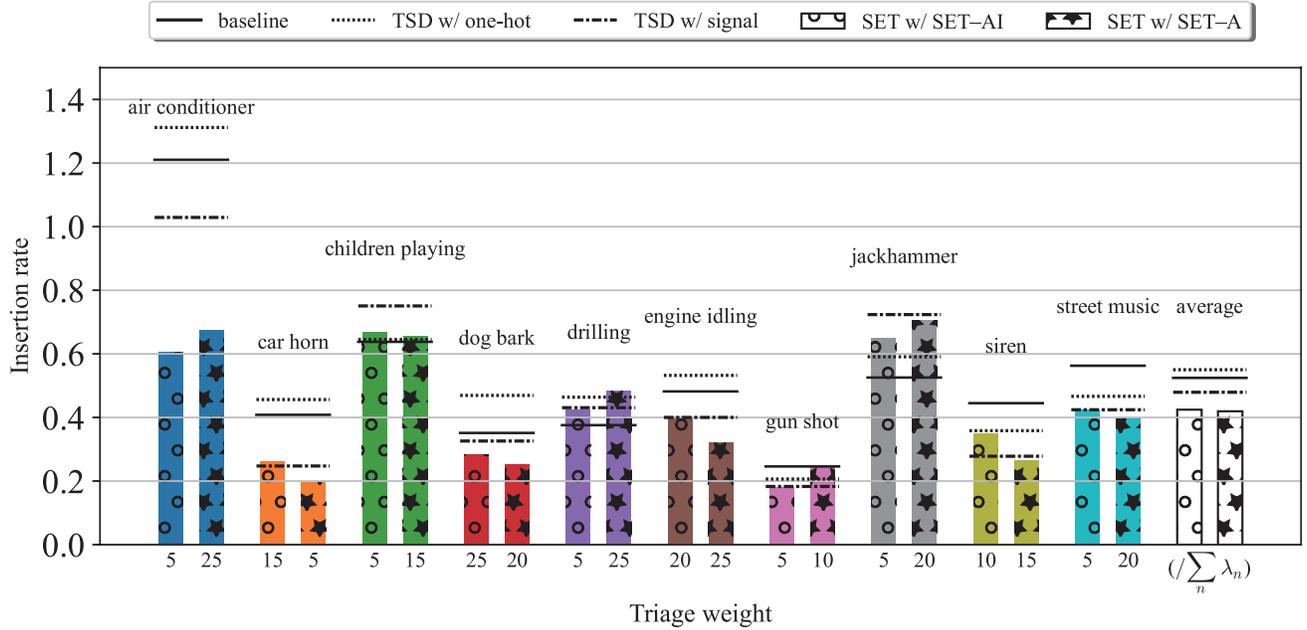}
\caption{SET results in terms of insertion rate for target classes}
\label{fig:result_03}
\end{figure*}
\vspace{15pt}
\begin{figure*}[h!]
\includegraphics[width=2.0\columnwidth]{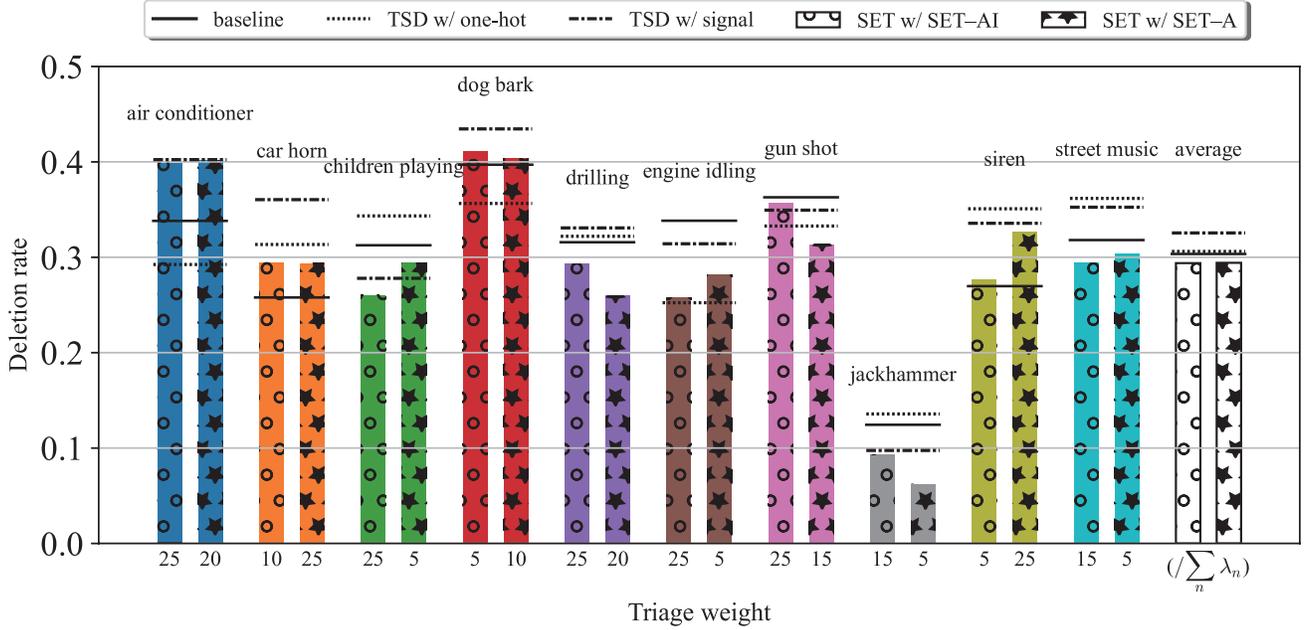}
\caption{SET results in terms of deletion rate for target classes}
\label{fig:result_04}
\end{figure*}

\subsubsection{\rm\textbf{[Experiment 2]}: SET results in terms of misdetection}
\vspace{5pt}
To investigate in more detail the performance of SET in Figs. \ref{fig:result_03} and \ref{fig:result_04}, we used the error-related evaluation metrics, that is, frame-based insertion rate (IR) and deletion rate (DR).

\begin{figure}[t!]
\centering
	\includegraphics[width=0.95\columnwidth]{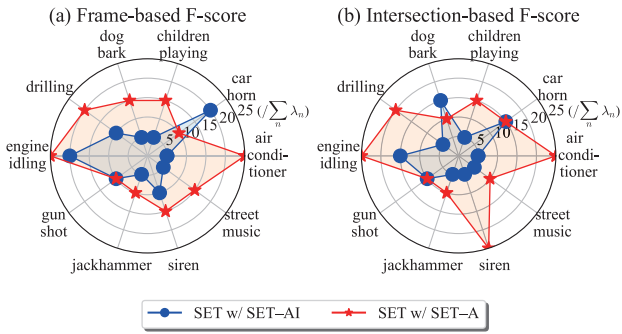}
	\caption{Optimal triage weights for each event class and evaluation metric}
	\label{fig:optimal_weight}
\end{figure}

\begin{figure}[t!]
\centering
\includegraphics[width=0.95\columnwidth]{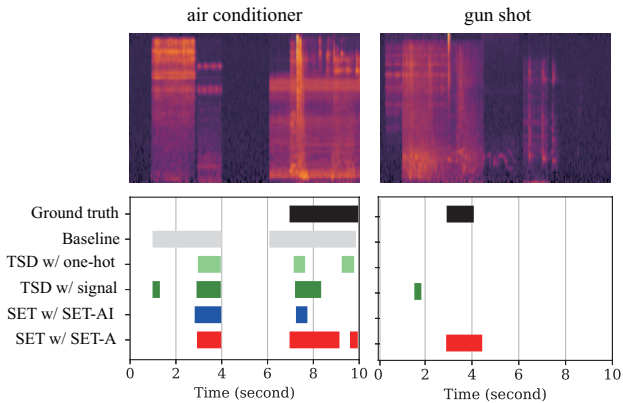}
\caption{Examples of detection results for each event}
\label{fig:example_detection}
\end{figure}
Given false positives (FPs) and false negatives (FNs) for each event and time frame $t$, IR and DR are defined using the insertion (I) and deletion (D) \cite{Mesaros_MDPI2016_01} as follows:

\begin{align}
    {\rm I}(n,t) &= {\rm max}(0, {\rm FP}(n,t)-{\rm FN}(n,t))\\[3pt]
    {\rm D}(n,t) &= {\rm max}(0, {\rm FN}(n,t)-{\rm FP}(n,t)) \\[3pt]
    {\rm IR}(n) &=  \frac{\sum_{t=1}^{T}{\rm I}(n,t)}{\sum_{t=1}^{T}{\rm A}(n,t)},\\[3pt]
    {\rm DR}(n) &=  \frac{\sum_{t=1}^{T}{\rm D}(n,t)}{\sum_{t=1}^{T}{\rm A}(n,t)},  
\label{eq:IR_DR}
\end{align}
\noindent where $n$ and $t$ represent the indexes of a sound event class and a time frame, respectively.
I($n,t$), D($n,t$), FP($n,t$), and FN($n,t$) are each a binary variable indicating whether there is I, D, FP, or FN of event $n$ at time $t$, respectively.
\textcolor{black}{A($n,t$) is a binary variable indicating whether event $n$ is active at frame $t$.}
IR($n$) and DR($n$) are the insertion and deletion rates for each event class $n$, respectively.

Figures \ref{fig:result_03} and \ref{fig:result_04} show results in terms of IR and DR for each target event.
The results show that the proposed SET methods outperformed the conventional methods in terms of  IR and DR.
In particular, SET w/ SET--AI \textcolor{black}{reduced} the IR of ``air conditioner'' by 0.605 points compared with the baseline value.
On the other hand, the IRs of the proposed SET for ``drilling'' and ``jackhammer'' are \textcolor{black}{higher} than the baseline values.
Our SET tends to increase false positives of acoustically similar classes more than the conventional method.
As shown in Figs. \ref{fig:result_03} and \ref{fig:result_04}, most of the classes suffer from the trade-off between IR and DR.
On the other hand, both the IR and DR of ``engine\_idling'' and ``street\_music'' are improved when using the proposed method compared with the baseline values.

Comparing the TSD and SET models, we find that the detection performance characteristics of the TSD models are unstable among the event classes compared with those of the SET models.
In other words, the difference in detection performance between SED and TSD is large and the superiority of one over the others often reverses compared with SED and SET.
In the training of the TSD models, the loss function of the target class is weighted with the same value even if the target class is being overtrained.
On the other hand, in our SET, the loss function of the target class is  weighted  with a random value referring to the Dirichlet distribution.
This random weighting may lessen the instability of the detection performance.

\subsubsection{\rm\textbf{[Experiment 3]}: Optimal triage weights of SET}
\vspace{5pt}
We then analyze the optimal triage weight for each event class and evaluation metric.
Figure \ref{fig:optimal_weight} shows the optimal triage weights represented by the radar chart.
As previously mentioned, the SET results are obtained using the optimal triage weight, which gives the best detection performance for each target event class, method, and evaluation metric.
As shown in Fig. \ref{fig:optimal_weight}, in most of the event classes, the optimal triage weights of SET w/ SET--A are larger than those of SET w/ SET--AI.
This is because the ability of detecting the active frames of the target class does not change significantly even when changing the triage weight with the SET--AI loss, as mentioned in Sec. 4.2.1.
In SET w/ SET--A, most of the optimal triage weights for the classes are between 10 and 20.
This is because the weight of 5 is very small for boosting the detection of the target class.
Weights over 25 are not densely sampled from the Dirichlet distribution we used.
In other words, the triage weight of the lower probability density does not greatly contribute to the training of the target class.
In practical applications, we can tune detection models for other evaluation metrics or scenes without retraining by the proposed methods.
This flexibility of the proposed methods has not been provided by the conventional methods.

Figure \ref{fig:example_detection} shows acoustic features and the system outputs and ground truths for selected events.
In this figure, the SET results where the triage weight is optimized for the intersection-based F-scores are selected.
In most of the cases, the proposed SET outperformed the conventional methods.
In particular, SET w/ SET--A is outstanding for detecting active frames for ``gun shot.''
In ``air conditioner,'' however, the proposed methods still produce false positives as with the conventional methods.
The problem of producing false positives needs to be solved in a future work.

%%%%%%%%%%%%%%%%%%%%%%%%
\section{Conclusion}
%%%%%%%%%%%%%%%%%%%%%%%%
In this work, we proposed a new task for SED: sound event triage (SET), in which an arbitrary number of event classes are prioritized.
In this first study of SET, we introduced training method for single target SET, which is a subtask of SET.
To perform single target SET, class-weighted training is used for detecting events with priority.
In the class-weighted training, loss functions and the network are stochastically weighted by the priority parameter of each class.
In inference stages, the single target SET network with class-weighted training can change into various specialists for each class without retraining.
Results of the experiments using the URBAN--SED dataset show that the proposed method with the SET--A loss outperforms the conventional SED method by 8.70, 6.66, and 6.09 percentage points for ``air\_conditioner,'' ``car\_horn,'' and ``street\_music,'' respectively, in terms of the intersection-based F-score.
The results revealed that the SET--A loss contributes more to the detection of a target class than the SET--AI loss.
In the average performance of the classes, the proposed methods increased the intersection-based F-score by 3.37 percentage points compared with the conventional SED and TSD models.
As the limitation of the proposed methods, the results indicate that the confusion errors among similar events might be enhanced.
In a future work, the multitarget SET needs to be studied by redesigning the distribution of the \textcolor{black}{priority used during training} for the multitarget SET.
Moreover, the SET performance for a small number of training data, e.g., one-shot or few-shot learning, also needs to be investigated.

\bibliographystyle{IEEEbib}
\bibliography{bmc_article}

\begin{thebibliography}{10}

\bibitem{Imoto_AST2018_01}
K.~Imoto,
\newblock ``Introduction to acoustic event and scene analysis,''
\newblock {\em Acoustical Science and Technology}, vol. 39, no. 3, pp.
  182--188, 2018.

\bibitem{koizumi_taslp2019_01}
Y.~Koizumi, S.~Saito, H.~Uematsum, Y.~Kawachi, and N.~Harada,
\newblock ``Unsupervised detection of anomalous sound based on deep learning
  and the {N}eyman-{P}earson lemma,''
\newblock {\em IEEE/ACM Trans. Audio, Speech, and Language Processing}, vol.
  27, no. 1, pp. 212--224, 2019.

\bibitem{Stork_ROMAN2012_01}
J.~Stork, L.~Spinello, J.~Silva, and {K. O. Arras},
\newblock ``Audio-based human activity recognition using non-{M}arkovian
  ensemble voting,''
\newblock in {\em Proc. {IEEE} International Symposium on Robot and Human
  Interactive Communication (RO-MAN)}, 2012, pp. 509--514.

\bibitem{Peng_ICME2009_01}
Y.-T. Peng, C.-Y. Lin, M.-T. Sun, and K.-C. Tsai,
\newblock ``Healthcare audio event classification using hidden {M}arkov models
  and hierarchical hidden {M}arkov models,''
\newblock in {\em Proc. {IEEE} International Conference on Multimedia and Expo
  (ICME)}, 2009, pp. 1218--1221.

\bibitem{Nandwana_INTERSPEECH2016_01}
M~Nandwana and T.~Hasan,
\newblock ``Towards smart-cars that can listen: Abnormal acoustic event
  detection on the road,''
\newblock in {\em Proc. {INTERSPEECH}}, 2016, pp. 2968--2971.

\bibitem{Ntalampiras_ICASSP2009_01}
S.~Ntalampiras, I.~Potamitis, and N.~Fakotakis,
\newblock ``On acoustic surveillance of hazardous situations,''
\newblock in {\em Proc. {IEEE} International Conference on Acoustics, Speech
  and Signal Processing (ICASSP)}, 2009, pp. 165--168.

\bibitem{Mesaros_SPmaga2021_01}
A.~Mesaros, T.~Heittola, T.~Virtanen, and M.~D. Plumbley,
\newblock ``Sound event detection: A tutorial,''
\newblock {\em {IEEE} Signal Processing Magazine}, vol. 38, no. 5, pp. 67--83,
  2021.

\bibitem{Mesaros_eusipco2010_01}
A.~{Mesaros}, T.~{Heittola}, A.~{Eronen}, and T.~{Virtanen},
\newblock ``Acoustic event detection in real life recordings,''
\newblock in {\em Proc. European Signal Processing Conference (EUSIPCO)}, 2010,
  pp. 1267--1271.

\bibitem{Heittola_JASM2013_01}
T.~Heittola, A.~Mesaros, A.~Eronen, and T.~Virtanen,
\newblock ``Context-dependent sound event detection,''
\newblock {\em EURASIP Journal on Audio, Speech, and Music Processing}, vol.
  2013, no. 1, pp. 1--13, 2013.

\bibitem{Gemmeke_WASPAA2013_01}
J.~Gemmeke, L.~{Vuegen}, P.~{Karsmakers}, B.~{Vanrumste}, and H.~{Van hamme},
\newblock ``An exemplar-based {NMF} approach to audio event detection,''
\newblock in {\em Proc. {IEEE} Workshop on Applications of Signal Processing to
  Audio and Acoustics (WASPAA)}, 2013, pp. 1--4.

\bibitem{Komatsu_DCASE2016_01}
T.~Komatsu, T.~Toizumi, R.~Kondo, and Y.~Senda,
\newblock ``Acoustic event detection method using semi-supervised non-negative
  matrix factorization with mixtures of local dictionaries,''
\newblock in {\em Proc. Workshop on Detection and Classification of Acoustic
  Scenes and Events (DCASE)}, 2016, pp. 45--49.

\bibitem{Hershey_ICASSP2017_01}
S.~Hershey, S.~Chaudhuri, D.~Ellis, J.~Gemmeke, A.~Jansen, C.~Moore, M.~Plakal,
  D.~Platt, R.~Saurous, B.~Seybold, M.~Slaney, R.~Weiss, and K.~Wilson,
\newblock ``{CNN} architectures for large-scale audio classification,''
\newblock in {\em Proc. {IEEE} International Conference on Acoustics, Speech
  and Signal Processing (ICASSP)}, 2017, pp. 131--135.

\bibitem{Hayashi_TASLP2017_01}
T.~Hayashi, S.~Watanabe, T.~Toda, T.~Hori, {J. Le Roux}, and K.~Takeda,
\newblock ``Duration-controlled {LSTM} for polyphonic sound event detection,''
\newblock {\em IEEE/ACM Trans. Audio, Speech, and Language Processing}, vol.
  25, no. 11, pp. 2059--2070, 2017.

\bibitem{cakir_TASLP2017_01}
E.~{\c{C}ak{\i}r}, G.~{Parascandolo}, T.~{Heittola}, H.~{Huttunen}, and
  T.~{Virtanen},
\newblock ``Convolutional recurrent neural networks for polyphonic sound event
  detection,''
\newblock {\em IEEE/ACM Trans. Audio, Speech, and Language Processing}, vol.
  25, no. 6, pp. 1291--1303, 2017.

\bibitem{kong_TASLP2020_01}
Q.~Kong, Y.~Xu, W.~Wang, and M.~D. Plumbley,
\newblock ``Sound event detection of weakly labelled data with
  {CNN}-{T}ransformer and automatic threshold optimization,''
\newblock {\em IEEE/ACM Trans. Audio, Speech, and Language Processing}, vol.
  28, pp. 2450--2460, 2020.

\bibitem{miyazaki_icassp2020_01}
K.~Miyazaki, T.~Komatsu, T.~Hayashi, S.~Watanabe, T.~Toda, and K.~Takeda,
\newblock ``Weakly-supervised sound event detection with self-attention,''
\newblock in {\em Proc. {IEEE} International Conference on Acoustics, Speech
  and Signal Processing (ICASSP)}, 2020, pp. 66--70.

\bibitem{miyazaki_DCASEc2020_01}
K.~Miyazaki, T.~Komatsu, T.~Hayashi, S.~Watanabe, T.~Toda, and K.~Takeda,
\newblock ``Conformer-based sound event detection with semi-supervised learning
  and data augmentation,''
\newblock in {\em Tech. Rep. DCASE Challenge}, 2020, pp. 1--5.

\bibitem{Slizovskaia_arxiv2022_01}
O.~Slizovskaia, G.~Wichern, Z.~Wang, and J.~L. Roux,
\newblock ``Locate this, not that: Class-conditioned sound event doa
  estimation,''
\newblock in {\em arXiv, arXiv:2203.04197}, 2022, pp. 1--5.

\bibitem{Yang_arxiv2021_01}
D.~Yang, H.~Wang, Y.~Zou, and C.~Weng,
\newblock ``Detect what you want: Target sound detection,''
\newblock in {\em arXiv, arXiv:2112.10153}, 2021, pp. 1--5.

\bibitem{Dosovitskiy_iclr2020_01}
A.~Dosovitskiy and J.~Djolonga,
\newblock ``You only train once: Loss-conditional training of deep net works,''
\newblock in {\em Proc. International Conference on Learning Representations
  (ICLR)}, 2020, pp. 1--17.

\bibitem{Perez_aaai2018_01}
E.~Perez, F.~Strub, H.~Vries, V.~Dumoulin, and A.~Courville,
\newblock ``{FiLM}: Visual reasoning with a general conditioning layer,''
\newblock in {\em Proc. the Association for the Advancement of Artificial
  Intelligence (AAAI)}, 2018, pp. 1--17.

\bibitem{imoto_icassp2021_01}
K.~Imoto, S.~Mishima, Y.~Arai, and R.~Kondo,
\newblock ``Impact of sound duration and inactive frames on sound event
  detection performance,''
\newblock in {\em Proc. {IEEE} International Conference on Acoustics, Speech
  and Signal Processing (ICASSP)}, 2021, pp. 860--864.

\bibitem{Kingma_ICLR2015_01}
D.~P. Kingma and J.~Ba,
\newblock ``Adam: A method for stochastic optimization,''
\newblock in {\em Proc. International Conference on Learning Representations
  (ICLR)}, 2015.

\bibitem{Salamon_WASPAA2017_01}
J.~Salamon, D.~MacConnell, M.~Cartwright, P.~Li, and J.~P. Bello,
\newblock ``Scaper: A library for soundscape synthesis and augmentation,''
\newblock in {\em Proc. {IEEE} Workshop on Applications of Signal Processing to
  Audio and Acoustics (WASPAA)}, 2017, pp. 344--348.

\bibitem{Serizel_DCASE2018_01}
R.~Serizel, N.~Turpault, H.~Eghbal-Zadeh, and A.~P. Shah,
\newblock ``Large-scale weakly labeled semi-supervised sound event detection in
  domestic environments,''
\newblock in {\em Proc. Workshop on Detection and Classification of Acoustic
  Scenes and Events (DCASE)}, 2018, pp. 19--23.

\bibitem{Bilen_ICASSP2020_01}
{Bilen \c{C}.}, G.~Ferroni, F.~Tuveri, J.~Azcarreta, and S.~Krstulovi\'e,
\newblock ``A framework for the robust evaluation of sound event detection,''
\newblock in {\em Proc. {IEEE} International Conference on Acoustics, Speech
  and Signal Processing (ICASSP)}, 2020, pp. 61--65.

\bibitem{Zheng_ICASSP2021_01}
X.~Zheng, Y.~Song, I.~McLoughlin, L.~Liu, and L.~Dai,
\newblock ``An improved mean teacher based method for large scale weakly
  labeled semi-supervised sound event detection,''
\newblock in {\em Proc. {IEEE} International Conference on Acoustics, Speech
  and Signal Processing (ICASSP)}, 2021, pp. 356--360.

\bibitem{Zheng_DCASEc2021_01}
X.~Zheng, H.~Chen, and Y.~Song,
\newblock ``Zheng {USTC} team's submission for {DCASE}2021 task4 -
  semi-supervised sound event detection,''
\newblock in {\em Tech. Rep. {DCASE} Challenge}, 2021, pp. 1--3.

\bibitem{Salamon_ACMMM14_01}
J.~Salamon, C.~Jacoby, and J.~P. Bello,
\newblock ``A dataset and taxonomy for urban sound research,''
\newblock in {\em Proc. 22nd {ACM} International Conference on Multimedia
  (ACM-MM'14)}, 2014, pp. 1041--1044.

\bibitem{Mesaros_MDPI2016_01}
A.~Mesaros, T.~Heittola, and T.~Virtanen,
\newblock ``Metrics for polyphonic sound event detection,''
\newblock {\em Applied Sciences}, vol. 6, no. 6, pp. 1--17, 2016.

\end{thebibliography}

\end{document}